\documentclass{emulateapj}

\def\kms{km~s$^{-1}$}
\def\degpnt{^{\circ}\kern-1.7mm.\kern+.35mm}
\def\arcpnt{"\kern-1.7mm.\kern+.35mm}
\def\minpnt{'\kern-1.0mm.\kern+.30mm}

\newcommand{\bvec}[1]{\mathbf #1}

\shorttitle{Peculiar Velocities of SNe}
\shortauthors{Neill, Hudson, \& Conley}

\begin{document}


\title{The Peculiar Velocities of Local Type I\lowercase{a} Supernovae and 
  their Impact on Cosmology}


\author{James~D.~Neill}
\affil{California Institute of Technology, 1200 E. California Blvd.,
  Pasadena, CA  91125}
\email{neill@srl.caltech.edu}

\author{Michael~J.~Hudson}
\affil{University of Waterloo, 200 University Avenue West, Waterloo, ON, 
N2L 3G1, CANADA}
\email{mjhudson@uwaterloo.ca}

\and

\author{Alex~Conley}
\affil{University of Toronto, 60 Saint George Street, Toronto, ON M5S 3H8, 
CANADA}
\email{conley@astro.utoronto.ca}


\begin{abstract}

  We quantify the effect of supernova Type Ia peculiar velocities on the
  derivation of cosmological parameters.  The published distant and local
  Ia SNe used for the Supernova Legacy Survey first-year cosmology report
  form the sample for this study.  While previous work has assumed that the
  local SNe are at rest in the CMB frame (the No Flow assumption), we test
  this assumption by applying peculiar velocity corrections to the local
  SNe using three different flow models.  The models are based on the IRAS
  PSCz galaxy redshift survey, have varying $\beta = \Omega_m^{0.6}/b$, and
  reproduce the Local Group motion in the CMB frame.  These datasets are
  then fit for $w$, $\Omega_m$, and $\Omega_\Lambda$ using flatness or
  $\Lambda$CDM and a BAO prior.  The $\chi^2$ statistic is used to
  examine the effect of the velocity corrections on the quality of the
  fits.  The most favored model is the $\beta=0.5$ model, which produces a
  fit significantly better than the No Flow assumption, consistent with
  previous peculiar velocity studies.  By comparing the No Flow assumption
  with the favored models we derive the largest potential systematic error
  in $w$ caused by ignoring peculiar velocities to be $\Delta w = +0.04$.
  For $\Omega_\Lambda$, the potential error is $\Delta \Omega_\Lambda =
  -0.04$ and for $\Omega_m$, the potential error is $\Delta \Omega_m <
  +0.01$.  The favored flow model ($\beta=0.5$) produces the following
  cosmological parameters: $w = -1.08^{+0.09}_{-0.08}$, $\Omega_m = 
  0.27^{+0.02}_{-0.02}$ assuming a flat cosmology, and $\Omega_\Lambda =
  0.80^{+0.08}_{-0.07}$ and $\Omega_m = 0.27^{+0.02}_{-0.02}$ for a
  $w = -1$ ($\Lambda$CDM) cosmology.

\end{abstract}




\section{Introduction}
\label{sec_intro}

Dark Energy has challenged our knowledge of fundamental physics since the
direct evidence for its existence was discovered using Type Ia supernovae
\citep{Riess98AJ, Perlmutter99ApJ}.  Because there are currently no
compelling theoretical explanations for Dark Energy, the correct emphasis,
as pointed out by the Dark Energy Task Force \citep[DETF,][]{Albrecht06},
is on refining our observations of the accelerated expansion of the
universe.  Recommendation V from the DETF Report \citep{Albrecht06} calls
for an exploration of the systematic effects that could impair the needed
observational refinements. 

A couple of recent studies \citep{Hui06PhRvD, Cooray06PhRvD} point out
that the redshift lever arm needed to accurately measure the universal
expansion requires the use of a local sample, but that coherent
large-scale local ($z < 0.2$) peculiar velocities add additional
uncertainty to the Hubble diagram and hence to the derived
cosmological parameters. 

Current analyses
\citep[e.g.,][]{Astier06A&A,Riess07ApJ,Wood-Vasey07astroph} of the
cosmological parameters do not attempt to correct for the effect of local
peculiar velocities.  As briefly noted by \citet{Hui06PhRvD} and
\citet{Cooray06PhRvD}, it is possible to use local data to measure the
local velocity field and hence limit the impact on the derived cosmological
parameters.  Measurements of the local velocity field have improved to the
point where there is consistency among surveys and methods \citep{Hud03,
HudSmiLuc04, Radburn04MNRAS, PikHud05, Sarkar06}.  Type Ia supernova
peculiar velocities have been studied recently by \cite{Radburn04MNRAS,
PikHud05, JhaRieKir06, HauHanTho06,Watkins07astroph} and others.  Their
results demonstrate that the local flows derived from SNe are in agreement
with those derived from other distance indicators, such as the Tully-Fisher
relation and the Fundamental Plane.  Our aim is to use the current
knowledge of the local peculiar motions to correct local SNe and, together with
a homogeneous set of distant SNe, fit for cosmological parameters and
measure the effect of the corrections on the cosmological fits. 

To produce this measurement, we analyze the local and distant SN~Ia sample
used in the first-year cosmology results from the Supernova Legacy Survey
\citep[SNLS,][hence A06]{Astier06A&A}.  This sample is composed of 44 local
SNe \citep[A06, Table~8:][]{Hamuy96AJ, Riess99AJ, Krisciunas01AJ,
Jha02PhDT, Strolger02AJ, Altavilla04MNRAS, Krisciunas04AJa,
Krisciunas04AJb} and 71 distant SNe (A06, Table~9).  The distant SNe are
the largest homogeneous set currently in the literature.  The local sample
span the redshift range $0.015 < z < 0.125$ and were selected to have good
lightcurve sampling (A06, \S~5.2).  Using three different models
encompassing the range of plausible local large-scale flow, we assign and
correct for the peculiar velocity of each local SN.  We then re-fit the
entire sample for $w$, $\Omega_m$, and $\Omega_\Lambda$ to assess the
systematics due to the peculiar velocity field, and to asses the change in
the quality of the resulting fits. 

\section{Peculiar Velocity Models}
\label{pecvel}

Peculiar velocities, $\bvec{v}$, arise due to inhomogeneities in the
mass density and hence in the expansion. Their effect is to perturb
the observed redshifts from their cosmological values: $cz_{\mathrm
  CMB} = cz + \bvec{v} \cdot \hat{\bvec{r}}$, where $cz$ is the
cosmological redshift the SN would have in the absence of peculiar
velocities.  With the advent of all-sky galaxy redshift surveys, it is
possible to predict peculiar velocities from the galaxy distribution
provided one knows $\beta = f(\Omega)/b$, where $b$ is a linear
biasing parameter relating fluctuations in the galaxy density,
$\delta$, to fluctuations in the mass density. The peculiar velocity
in the CMB frame is then given by linear perturbation theory
\citep{Peebles80book} applied to the density field \citep[see,
e.g.][]{YahStrDav91,Hud93}:
\begin{equation}
	\label{eq_pvel}
\bvec{v} = \frac{\beta}{4\pi} \int^{R_{max}} \delta(\bvec{r}')
\frac{(\bvec{r}'-\bvec{r})}{|\bvec{r}'-\bvec{r}|^3} d^3 \bvec{r'} +
\bvec{V}. 
\end{equation}

In this Letter, we use the density field of IRAS PSCz galaxies
\citep{BraTeoFre99}, which extends to a depth $R_{max}=20000$ \kms. 
Contributions to the peculiar velocity arising from masses on scales
larger than $R_{max}$ are modeled by a simple residual dipole,
$\bvec{V}$.  Thus, given a density field, the parameters $\beta$ and
$\bvec{V}$ describe the velocity field within $R_{max}$.  For galaxies
with distances greater than $R_{max}$, the first term above is set to
zero.

The predicted peculiar velocities from the PSCz density field are subject
to two sources of uncertainty: the noisiness of the predictions due to the
sparsely-sampled density field, and the inapplicability of linear
perturbation theory on small scales.  Typically these uncertainties are
accounted for by adding an additional ``thermal'' dispersion, which is
assumed to be Gaussian. From a careful analysis of predicted and observed
peculiar velocities, \citet{Willick98ApJ} estimated these uncertainties to
be $\sim 100$ \kms,  albeit with a dependence on density.
\citet{Radburn04MNRAS} found reasonable $\chi^2$ values if 150 \kms\ was
assumed in the field, with an extra contribution to the small-scale
dispersion added in quadrature for SNe in clusters. Here we adopt a thermal
dispersion of 150 \kms.

For this study, we explore the results of three different models of
large-scale flows and compare them to a case where no flow model is used.
These models have been chosen to span the range of flow models permitted by
peculiar velocity data, and all of these models reproduce the observed
$\sim$ 600 km s$^{-1}$ motion of the Local Group with respect to the CMB.
The first model assumes a pure bulk flow (model PBF, hence $\beta=0$) with
$\bvec{V}$ having vector components $(57,-540,314)$ km s$^{-1}$ in Galactic
Cartesian coordinates.  The second model assumes $\beta=0.5$ (model B05),
with a dipole vector of $(70,-194,0)$ km s$^{-1}$.  The third model adopts
$\beta=0.7$ (model B07) which requires no residual dipole.  We compare
these models to the no-correction scenario adopted by A06 and others with
$\beta=0$, $V=0$ which we call the ``No Flow'' or NF scenario.  Note that a
recent comparison \citep{PikHud05} of results from IRAS predictions versus
peculiar velocity data yields a mean value fit with $\beta=0.50\pm0.02$
(stat), so the B05 model is strongly favored over the NF scenario by
independent peculiar velocity analyses. 

\section{Cosmological Fits}
\label{sec_cosmo}

Prior to the fitting procedure, the peculiar velocities for each model are
used to correct the local SNe \citep[using a variation of][equations 11 and
13]{Hui06PhRvD}.  We then fit our corrected SN data in two ways using a
$\chi^2$-gridding cosmology fitter\footnote{\tt
http://qold.astro.utoronto.ca/conley/simple\_cosfitter/} \citep[also used
by][]{Wood-Vasey07astroph}.  The first fit uses a flat cosmology
($\Omega=1$) with the equation of state parameter $w$ and $\Omega_m$ as
free parameters.  The second fit assumes a $\Lambda CDM$ ($w=-1$) cosmology
with $\Omega_\Lambda$ and $\Omega_m$ as free parameters.  We used the same
intrinsic SN photometric scatter ($\sigma_{int} = 0.13$ mag, A06) for every
fit.  The resulting $\chi^2$ probability surfaces for both fits are then
further constrained using the BAO result from \citet{Eisenstein05ApJ}.  The
final derived cosmological parameters are then used to calculate the
$\chi^2$ for each fit (see A06, \S~5.4).

The fitting procedure employed here differs in implementation from that
used in A06.  Three additional parameters, often called nuisance
parameters, must be fit along with the two cosmological parameters.  These
parameters are the constant of proportionality for the SN lightcurve shape,
$\alpha_s$, the correction for the SN observed color, $\beta_c$, and a SN
brightness normalization, $\mathcal{M}$.  We distinguish $\beta_c$ from the
$\beta$ used to describe the flow models above.  A06 used analytic
marginalization of the nuisance parameters $\alpha_s$ and $\beta_c$ in
their fits.  Here these parameters are fully gridded like the cosmological
parameters.  This avoids a bias in the nuisance parameters that results
because, in the analytic method, their values must be held fixed to compute
the errors.  The result is that our fits using the NF scenario produces
slightly different cosmological parameters than quoted in A06.  

\section{Results}

\begin{deluxetable*}{lccccccccc}
	\tablecaption{Peculiar Velocity Model Parameters and Results\label{tab_models}}
	\tablewidth{0pt}
	\tablehead{
	&  & & \multicolumn{3}{c}{$\Omega = 1$ + BAO prior} & &
	\multicolumn{3}{c}{$w = -1$ + BAO prior} \\
	\cline{4-6} \cline{8-10} \\
	\colhead{Model} & \colhead{$\beta$} & \colhead{$\bvec{V}$ (\kms )} & 
	\colhead{$w$} & \colhead{$\Omega_m$} & 
	\colhead{$\chi^2_{w,\Omega_m}$} & &
	\colhead{$\Omega_\Lambda$} & \colhead{$\Omega_m$} &
	\colhead{$\chi^2_{\Omega_\Lambda,\Omega_m}$} \\
	}

\startdata
A06\tablenotemark{a}  & 0.0 & \nodata     & 
$-1.023\pm0.090$ & $0.271\pm0.021$ & \nodata & &
$0.751\pm0.082$ & $0.271\pm0.020$ & \nodata \\[2pt]
NF  & 0.0 & \nodata     & 
$-1.054^{+0.086}_{-0.084}$ & $0.270^{+0.024}_{-0.018}$ & 115.5 & &
$0.770^{+0.083}_{-0.071}$ & $0.269^{+0.033}_{-0.017}$ & 115.4 \\[2pt]
PBF & 0.0 & 57,-540,314 & 
$-1.026^{+0.085}_{-0.083}$ & $0.273^{+0.024}_{-0.019}$ & 129.4 & &
$0.741^{+0.084}_{-0.073}$ &  $0.273^{+0.034}_{-0.017}$ & 129.2 \\[2pt]
B05 & 0.5\tablenotemark{b} & 70,-194,0   & 
$-1.081^{+0.087}_{-0.085}$ & $0.268^{+0.024}_{-0.018}$ & 110.3 & &
$0.796^{+0.081}_{-0.070}$ &  $0.267^{+0.032}_{-0.017}$ & 110.1 \\[2pt]
B07 & 0.7 & \nodata     & 
$-1.094^{+0.087}_{-0.085}$ & $0.267^{+0.024}_{-0.018}$ & 111.2 & &
$0.809^{+0.082}_{-0.069}$ & $0.265^{+0.032}_{-0.017}$ & 111.1 \\
\enddata
\tablenotetext{a}{results quoted in A06 marginalizing
analytically over $\alpha_s$ and $\beta_c$ (see \S~\ref{sec_cosmo})}
\tablenotetext{b}{best fit value from \citet{PikHud05}}
\end{deluxetable*}

\begin{figure}
\includegraphics[scale=0.35,angle=90.,viewport=0 0 500 700]{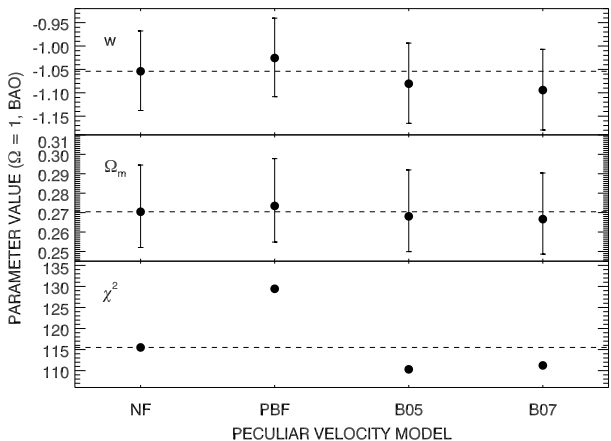}
\caption{Parameter values for the $w$, $\Omega_m$ fit ($\Omega=1$
+ BAO prior) for each of the four peculiar velocity models in
Table~\ref{tab_models}.  The values for the NF scenario are indicted by the
dashed lines.  The largest systematic error in $w$ compared 
with the NF fit is $+0.040$ for the B07 model, which
demonstrates the amplitude of the systematic error if peculiar velocity is
not accounted for.  The offsets for $\Omega_m$ are all within $\pm 0.003$
showing that this parameter is not sensitive to the peculiar velocity
corrections due to the BAO prior.  The $\chi^2$ of the fits improve when 
using the two $\beta$ models (B05, B07), while the PBF model provides a 
significantly worse fit. 
}
\label{fig_wom_fit}
\end{figure}

\begin{figure}
\includegraphics[scale=0.35,angle=90.,viewport=0 0 500 700]{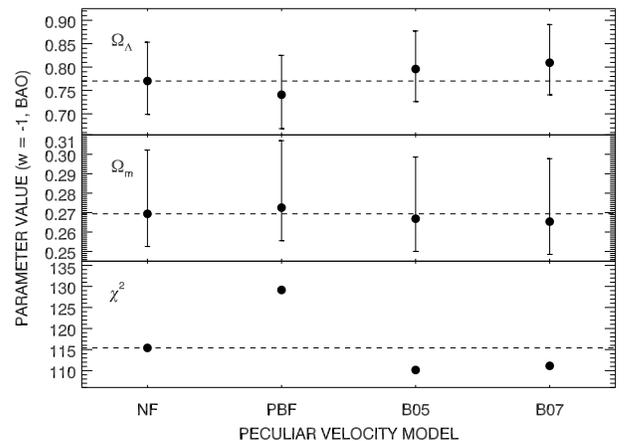}
\caption{Parameter values for the $\Omega_\Lambda$, $\Omega_m$ fit ($w = 
-1$ + BAO prior) for each of the four peculiar velocity models as in
Figure~\ref{fig_wom_fit}.  Again, comparing the NF fits to the B07 
model produces the largest systematic in $\Omega_\Lambda$ of $-0.039$. 
We also find $\Omega_m$ insensitive to the corrections, having 
all offsets within $\pm 0.004$.  The $\chi^2$ values show the same pattern 
as in Figure~\ref{fig_wom_fit}, favoring the $\beta$ models over no 
correction (NF), and over pure bulk flow. 
}
\label{fig_omol_fit}
\end{figure}

The results of the cosmological fits for each model are listed in
Table~\ref{tab_models} and plotted in Figure~\ref{fig_wom_fit} and
Figure~\ref{fig_omol_fit}.  They demonstrate two effects of the
peculiar velocity corrections: a change in the values of the
cosmological parameters, and a change in the quality of the fits as
measured by the $\chi^2$ statistic. 

We expect, if a given model is correct, to improve the fitting since our
corrected data should more closely resemble the homogeneous universe
described by a few cosmological parameters.  The $\chi^2$ of the fits for
each flow model can be compared to the $\chi^2$ for the NF scenario (shown
by the dashed line in the figures) as a test of this hypothesis.  Using
$\Delta \chi^2 = -2 \ln L/L_{NF}$, where $L$ is the likelihood, we find
that the pure bulk flow is over 10$^3$ times less likely than the NF
scenario, while the B05 and B07 models are 13.5 and 8.6 times more likely,
respectively.

We also use these data to assess the systematic errors made in the
parameters if no peculiar velocities are accounted for.  The largest of
these are obtained by comparing the B07 model with the NF scenario.  This
comparison yields $\Delta w_{B07} = +0.040$ and $\Delta\Omega_{\Lambda,B07}
= -0.039$.  The same comparison for the B05 model, which is only slightly
preferred by the $\chi^2$ statistic over model B07, produces $\Delta
w_{B05} = +0.027$ and $\Delta\Omega_{\Lambda,B05} = -0.026$.  The
systematic offsets for $\Omega_m$ are all $0.004$ or less, demonstrating
the insensitivity of this parameter to peculiar velocities.  This is due to
the BAO prior which is insensitive to local flow and provides a much
stronger constraint for $\Omega_m$ than for $w$ or $\Omega_\Lambda$
(see A06, Figures~5 and 6).

\section{Discussion and Summary}

The systematic effect of different flow models is at the level of $\pm0.04$
in $w$. This is smaller than the \emph{present} level of random error in
$w$, which is largely due to the small numbers of high- and low-redshift
SNe.  However, compared to other systematics discussed in A06, which total
$\Delta w = \pm0.054$, the systematic effect of large-scale flows is
important.  \citet[Table 5]{Wood-Vasey07astroph} list 16 sources of
systematic error which total $\Delta w = \pm 0.13$.  Aside from three
method-dependent systematics and the photometric zero-point error, they
are all smaller than the flow systematic.  As the number of SNe continues
to increase, and understanding of other systematics (e.g. photometric
zero-points) improves, it is possible that large-scale flows will become
one of the dominant sources of systematic uncertainty. 

The peculiar velocities of SN host galaxies arise from large-scale
structures over a range of scales. The component arising from
small-scale, local structure is the least important: it is essentially
a random variable which is reduced by $\sqrt{N}$.  More problematic is
the large-scale coherent component.  Such a large-scale component can
take several forms: an overdensity or underdensity; a large-scale
dipole, or ``bulk'' flow. 

The existence of a large-scale, but local ($<7400$ \kms) underdensity, or
``Hubble Bubble'' was first discussed by \cite{ZehRieKir98}.  Recently
\cite{JhaRieKir06} have re-enforced this claim with a larger SN data set:
they find that the difference in the Hubble constant inside the Bubble and
outside is $\Delta H/H = 6.5\pm1.8\%$.  If correct, this could have a
dramatic effect on the derived cosmological parameters \citep[Fig
17]{JhaRieKir06}, especially for those studies that extend their local
sample down below $z<0.015$.  However, the ``Hubble Bubble'' was not
confirmed by \cite{GioDalHay99} who found $\Delta H/H = 1.0\pm2.2$\% using
the Tully-Fisher (TF) peculiar velocities, nor by \cite{HudSmiLuc04} who
found $\Delta H/H = 2.3\pm1.9\%$ using the Fundamental Plane (FP)
distances.

According to equation~\ref{eq_pvel}, a mean underdensity of IRAS galaxies
of order $\sim$ 40\% within 7400 \kms\ would be needed to generate the
``Hubble Bubble'' quoted by \cite{JhaRieKir06}.  However, we find that the
IRAS PSCz density field of \cite{BraTeoFre99} is not underdense in this
distance range; instead it is mildly overdense (by a few percent) within
7400 \kms\ \citep[see also][Figure 2]{BraTeoFre99}.  As a further
cross-check, when we refit the \cite{JhaRieKir06} data after having
subtracted the predictions of the B05 flow model, the ``Bubble'' remains in
the \cite{JhaRieKir06} data.  Thus, the Jha et al ``Bubble'' cannot be
explained by local structure, unless that structure is not traced by IRAS
galaxies.  Moreover, when we analyze the 99 SNe within 15000 \kms\ from
\cite{Tonry03ApJ} in the same way, we find no evidence of a significant
``Hubble Bubble'' ($\Delta H/H = 1.5\pm2.0$\%), in agreement with the results
from TF and FP surveys.  The \citet{Tonry03ApJ} sample and that of
\cite{JhaRieKir06} have 67 SNe in common.  The high degree of overlap
suggests that the difference lies in the different methods for converting
the photometry into SN distance moduli. 

A local large-scale flow can also introduce systematic errors if the
low-z sample is biased in its sky coverage: in this case, an
uncorrected dipole term can corrupt the monopole term, which then
biases the cosmological parameters.  For the large-scale flow directions
considered here, this does not appear to affect the A06 sample: we
note that the PBF-corrected case has similar cosmological parameters
to the ``No Flow'' case.  However, if coherent flows exist on
large scales, this may affect surveys with unbalanced sky coverage,
such as the SN Factory \citep{Aldering02SPIE} or the SDSS SN
survey\footnote{\tt http://sdssdp47.fnal.gov/sdsssn/sdsssn.html}. 

The most promising approach to treating the effect of large-scale flows is
a more sophisticated version of the analysis presented here: combine
low-redshift SNe with other low-redshift peculiar velocity tracers, such as
Tully-Fisher SFI++ survey \citep{MasSprHay06} and the NOAO Fundamental
Plane Survey \citep{SmiHudNel04}, and use these data to constrain the
parameters of the flow model ($\beta$ and the residual large-scale flow
$\bvec{V}$) directly.  One can then marginalize over the parameters of the
flow model while fitting the cosmological parameters to the low- and
high-$z$ SNe. 



\acknowledgments





\end{document}